\documentclass[a4paper,aps,pra,showpacs,preprintnumbers,twocolumn]{revtex4}
\usepackage[cp1251]{inputenc}
\usepackage[english]{babel}
\usepackage{amssymb,amsfonts,amsmath,mathtext,enumerate,float,dsfont}
\usepackage{graphics,graphicx,epsfig,epstopdf}
\usepackage{caption}
\usepackage{indentfirst}
\usepackage[usenames]{color}
\usepackage{amsthm}

\begin{document}
\title{Finding the optimal cluster state configuration. Minimization of one-way quantum computation errors.}
\author{Korolev S. B.}
\email{Sergey.Koroleev@gmail.com}
\author{Golubeva T. Yu.}
\author{Golubev Yu. M.}
\affiliation{St. Petersburg State University, Universitetskaya  nab. 7/9, 199034 St. Petersburg, Russia}
\date{\today}
\pacs{03.65.Ud, 03.67.Bg, 03.67.-a, 03.67.Lx }

\begin{abstract}
In this paper, we estimate the errors of Gaussian transformations implemented using one-way quantum computations on cluster states of various configurations. From all possible cluster state configurations, we choose those that give the smallest computation error.  Furthermore, we evaluate errors in hybrid computational schemes, in which Gaussian operations are performed using one-way computations with additional linear transformations. As a result, we find the optimal strategy for the implementation of universal Gaussian computations with minimal errors.
\end{abstract}
\maketitle

\section{Introduction}
One-way quantum computation (OWQC) is one of the promising models of quantum computation \cite{Raus2}. To perform transformations in this model, it is not necessary to create any additional devices that implement logical operations on quantum states. All transforms in OWQC are realized using local measurements of a multipartite entangled state called the cluster state \cite{Raussendorf_1}. By varying the bases of measuring instruments, one can perform certain quantum transformations over the input states.

By definition, a cluster state is a multipartite entangled state characterized by some undirected mathematical graph. The graph nodes are physical systems, and the edges are entanglements between them. It follows that for cluster state preparation, it is necessary to take physical systems and entangle them between each other so that the resulting state has a certain configuration (a certain graph). The results of OWQC depend on the configurations of the cluster state graphs. At first glance, it seems that the richer graph configuration we have, the wider our computational capabilities. One can imagine a wide variety of different configurations, so the question arises as to which configuration is best used for computation. In other words, is it possible to indicate some "optimal" cluster configuration that is best suited for a particular type of computing? To answer this question, first of all, we need to identify restrictions that can make us abandon configurations with a large number of edges. To do this, we turn to the description of physical systems on which cluster states can be generated.

As such physical systems, one can use both systems described by discrete and continuous variables (CV). In this work, we will be interested in variables of the second type. Any continuous variables physical system is described as a harmonic oscillator with two quadratures $ \hat{x}_s$ and $ \hat{y}_s$, which obey the canonical commutation relation $\left[\hat x_s, \hat y_s \right] = {i}/{2} $. The main condition for using such systems to generate cluster states is their squeezing (for certainty, we will assume that the oscillators are squeezed by the $\hat{y}_s$ quadrature). The squeezing of quadrature means that its variance is less than the vacuum state variance ($\langle \delta \hat y_s ^2\rangle \textless {1}/{4}$).

The main restriction in the OWQC with continuous variables is an error in the results of computations associated with the use of physical systems with finite squeezing. Each such system (each cluster state node) adds to the computation result an error proportional to its squeezing degree. Therefore, the computation error will increase with the increasing number of cluster state nodes. Thus, to error minimization, it is necessary to use cluster states with a minimum node number. On the other hand, the number of nodes used should be sufficient to implement quantum transformations. So we naturally come to the concept of the optimal node number.

We will also consider the CV states as input states over which the computations will be performed. It is sufficient to be able to perform two types of operations for implementation of any unitary transformations over CV states: arbitrary Gaussian and one non-Gaussian \cite{Lloyd}. Any Gaussian operation, in turn, can be implemented by sequential application of an arbitrary single-mode transformation, a two-mode  \emph{CZ} transformation, and a displacement transformation of the quadratures. In this paper, we will be interested in finding configurations of cluster states only for Gaussian operations. We will not consider non-Gaussian transforms since an optimal algorithm for OWQC using a two-node cluster state (a state with a minimum non-trivial number of nodes) has already been theoretically proposed for their implementation.

In paper \cite{Korolev_2020}, we found cluster state configurations on which universal Gaussian transformations can be performed. To do this, we divided all possible configurations into several groups, which differ from each other in the number of nodes and whether the input states (the states that we want to transform) mix with the measured or unmeasured cluster state modes. We have obtained explicit relations between input and output states for each of the configuration groups. This allowed us to determine the transformations that can be performed in each of these groups. As a result, we have identified two types of interesting configurations.  The first type includes configurations on which universal OWQC can be implemented directly, without any additions. The second type contains cluster configurations on which only some Clifford group generators can be realized. In this article, we supplement the second type configurations so that they can be used to perform universal Gaussian transformations. Further, we compare the errors in all available universal transformation schemes. As a result, we will find universal computing schemes that give the minimum error for any squeezing of the used quantum oscillators. We will also present a recipe according to which one can select a cluster and build a computational scheme that ensures the transformation with minimal error.
 
 \section{Unmodified one-way quantum computation}
We begin our analysis with the case of unmodified OWQC when a single cluster state is used for transforming input states without any additions. In this case, it is impractical to use cluster states with a large number of nodes to perform universal Gaussian transformations, since an error will accumulate in the results. As a result, there may be a situation where the error overlaps a useful result, and it will be impossible to compensate with quantum error correction codes \cite{QECC}. The preferred way to implement universal Gaussian computation is to sequentially transform input states using clusters of two configuration types. Cluster states with the first type configuration should be suitable for performing arbitrary single-mode operations, and clusters with the second type configuration should be suitable for the two-mode operation \emph{CZ}. In doing so, each configuration should have the smallest number of nodes to guarantee the minimum error of the corresponding transformation. In addition, the computations performed in this way can be alternated with the error correction procedure so that the error does not accumulate in the results. Thus, when performing universal Gaussian operations in this way, the resulting error will be the smallest. Let's get to evaluating this error.

\subsection{Single-mode transformations} \label{cluster_one_mode}
\begin{figure} [H]
\centering
\includegraphics[scale=1]{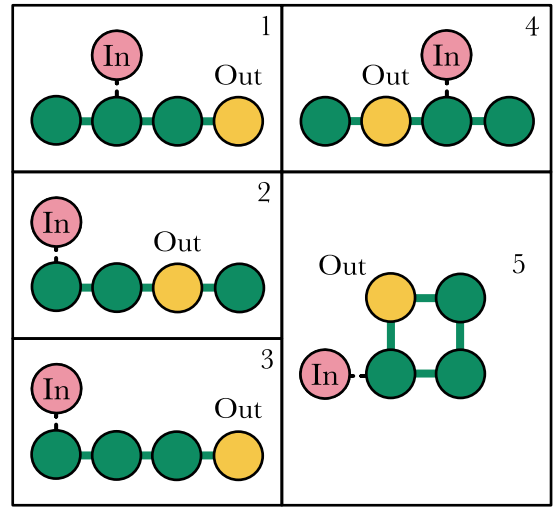}
\caption{Five types of cluster states configurations by which one can implement universal single-mode transformations: dashed lines indicate the input states In, which will mix with the nodes of the cluster state; Out is the unmeasured node of the cluster, the state in which the system will be after all the transformations} \label{Fig_1}
\end{figure}

First, let's estimate the error of universal single-mode transformations. In \cite{Korolev_2020,Ukai}, we have shown that the minimal number of cluster state nodes necessary to perform these transformations is four. We also have found that there are only two types of cluster configurations that satisfy the requirement of universality: linear and square (see Fig. \ref{Fig_1}). Four different types of universal single-mode transformations can be performed on a linear cluster state. This follows from possible mutual arrangements on the graph of the input node (the cluster node to which the input state is mixed) and the unmeasured (output) node. Note that these nodes are highlighted by the OWQC procedure itself. For a square cluster state, the situation is different, since this state can be used for universal single-mode calculations only if the input and output nodes are neighbors. This restriction and the symmetry of the square cluster state leads to the fact that only one type of transformations we need can be implemented in this state. All five ways to perform a universal single-mode Gaussian transformation are shown in Fig. \ref{Fig_1}.

Let us turn to the relations between the output and input ($\hat{x}_{in}, \hat{y}_{in}$) quadratures obtained by OWQC on the cluster states of the presented configurations (an explicit form of the formulas was obtained in \cite{Korolev_2020}):   
\begin{align} \label{tranform}
\begin{pmatrix}
\hat X_{out,j}\\
\hat Y_{out,j}
\end{pmatrix} = U_j \begin{pmatrix}
\hat{x}_{in}\\
\hat{y}_{in}
\end{pmatrix}+\hat{\vec e}_j, \quad j=1.\dots 5,
\end{align}
where subscript $j$ indicates the cluster configuration number, and the transformation matrices are given by the expressions:
\begin{align}
{U}_{1}&=\begin{pmatrix}
 -\cot \Theta _{4}^1 \tan \Theta _{3} ^1-1& \cot \Theta _{4}^1 \\
 \tan \Theta _{3}^1 & -1 \\
\end{pmatrix}R\left(\frac{\pi}{2}\right) \nonumber\\
&*R\left( -\frac{1}{2}\Theta_+^1\right)S\left(\ln \left[\tan \frac{1}{2}\Theta_-^1\right]\right)R\left(-\frac{1}{2}\Theta_+^1\right), \label{1}\\
{U}_{2}&=R\left(\frac{\pi}{2} \right)\begin{pmatrix}
 -\cot \Theta _{3}^2 \tan\Theta _{4}^2 -1& -\tan \Theta
   _{4}^2 \\
 -\cot \Theta _{3}^2 & -1 \\
\end{pmatrix}\nonumber\\
&*R\left(-\frac{1}{2}\Theta_+^2\right)S\left(\ln \left[\tan \frac{1}{2}\Theta_-^2\right]\right)R\left(-\frac{1}{2}\Theta_+^2\right),
\end{align}
\begin{align}
{U}_{3}&=\begin{pmatrix}
\cot\Theta_{4}^3 \cot \Theta_{3}^3-1  &  \cot \Theta_{4}^3\\
-\cot \Theta_{3}^3 & -1
\end{pmatrix}R\left(\pi \right)\nonumber\\
 &*R\left(-\frac{1}{2}\Theta_+^3\right)S\left(\ln \left[\tan \frac{1}{2}\Theta_-^3\right]\right)R\left(-\frac{1}{2}\Theta_+^3\right), \label{one_3}\\
{U}_{4}&=R\left(-\frac{\pi}{2} \right)\begin{pmatrix}
\tan \Theta _{3}^4\tan \Theta _{4}^4-1 & -\tan \Theta _{4}^4 \\
 \tan \Theta _{3}^4 & -1 
\end{pmatrix}R\left(\frac{\pi}{2}\right)\nonumber\\
&*R\left(-\frac{1}{2}\Theta_+^4\right)S\left(\ln \left[\tan \frac{1}{2}\Theta_-^4\right]\right)R\left(-\frac{1}{2}\Theta_+^4\right), 
\end{align}
\begin{align}
{U}_{5}&=\begin{pmatrix}
 \tan \Theta_{ 3} ^5\tan \Theta_{4}^5-1 & -\tan \Theta_{4}^5 \\
 \tan \Theta_{ 3}^5 & -1 
\end{pmatrix}\nonumber\\
&*R\left(-\frac{1}{2}\Theta_+^5\right)S\left(\ln \left[\tan \frac{1}{2}\Theta_-^5\right]\right)R\left(-\frac{1}{2}\Theta_+^5\right) \label{5},
\end{align}
and vectors of the computation errors, depending on the squeezing of quantum oscillators used in the computation, have the following form:
\begin{widetext}
\begin{align} 
&\hat{\vec{e}}_1=\frac{1}{d_2}\begin{pmatrix}
 3 \cot \Theta _{4}^1 & \cot \Theta _{4}^1 & -1-2 \cot
  \Theta _{4}^1 \tan \Theta _{3}^1 & -3-\cot \Theta
   _{4} ^1\tan \Theta _{3}^1 \\
 -2 & 1 & 2 \tan \Theta _{3}^1 & \tan \Theta _{3}^1
\end{pmatrix}
\begin{pmatrix}
 d_1-1 & -{1} & 0 & 0 \\
 -1 & d_1 & 0 & 0 \\
 0 & 0 & d_1 & -1 \\
 0 & 0 & -1  & d_1-1
\end{pmatrix} \hat{\vec{y}}_s, \label{er_1}\\
&\hat{\vec{e}}_2=\frac{1}{d_2}\begin{pmatrix}
 2 \cot  \Theta _{3}^2  & \cot  \Theta _{3}^2  & -3 & -1 \\
 -2 \cot  \Theta _{3}^2  \tan  \Theta _{4}^2 -1 & 2-\cot  \Theta
   _{3} ^2 \tan  \Theta _{4}^2  & 2 \tan  \Theta _{4}^2  &
   -\tan \Theta _{4}^2
\end{pmatrix}\begin{pmatrix}
 d_1 & -{1} & 0 & 0 \\
 -1 & d_1-1 & 0 & 0 \\
 0 & 0 & d_1-1 & -1 \\
 0 & 0 & -1  & d_1
\end{pmatrix}\hat{\vec{y}}_s,\\
&\hat{\vec{e}}_3=\frac{1}{d_2}\begin{pmatrix}
1-2\cot \Theta_{3}^3 \cot \Theta_{4}^3 & \cot \Theta_{3}^3 & 3\cot \Theta_{3}^3 & -2-\cot \Theta_{3}^3 \cot \Theta_{4}^3 \\
2\cot \Theta_{4}^3 & 1 & -2 & \cot \Theta_{4}^3 
\end{pmatrix}\begin{pmatrix}
 d_1 & 0 & 0 & -1 \\
 0 & d_1 & -1 & 0 \\
 0 & -1 & d_1-1 & 0 \\
 -1 & 0 & 0  & d_1-1
\end{pmatrix}\hat{\vec{y}}_s,\\
&\hat{\vec{e}}_4=\frac{1}{d_2}\begin{pmatrix}
 -3 & \tan  \Theta _{3}^4  & 2 \tan  \Theta _{3}^4  & -1 \\
 2 \tan  \Theta _{4}^4  & 3-\tan  \Theta _{3}^4  \tan  \Theta
   _{4}^4  & 1-2 \tan  \Theta _{3}^4  \tan  \Theta _{4}^4  &
   -\tan \Theta _{4} 
\end{pmatrix}\begin{pmatrix}
d_1-1 & 0 & 0 & -1 \\
 0 & d_1-1 & -1 & 0 \\
 0 & -1 & d_1 & 0 \\
 -1 & 0 & 0  & d_1 
\end{pmatrix}\hat{\vec{y}}_s,\\
&\hat{\vec{e}}_5=\frac{1}{d_1+2}\begin{pmatrix}
 \tan \Theta_{ 3} ^5\tan \Theta_{ 4}^5-3 & 2 \tan \Theta_{ 4}^5 & 3 \tan \Theta_{ 4}^5 & -\tan \Theta_{ 3}^5 \tan
   \Theta_{ 4}^5-2  \\
 \tan \Theta_{ 3}^5 & 3  & 2& -\tan \Theta_{3}^5 
\end{pmatrix}\begin{pmatrix}
 d_1 & 0 & 0 & -2 \\
 0 & d_1 & -2 & 0 \\
 0 & -2 & d_1 & 0 \\
 -2 & 0 & 0  & d_1
\end{pmatrix}\hat{\vec{y}}_s. \label{er_5}
\end{align}
\end{widetext}
In the equations above, we introduced the following notation: $d_1=\sqrt{5}+3$; $d_2=\sqrt{5\left(5+2\sqrt{5}\right)}$; $\Theta_{\pm}^j=\Theta_{2}^j\pm \Theta_{1}^j$; $ \hat{\vec y}_s=\begin{pmatrix} \hat{y}_{s,1},\hat{y}_{s,2},\hat{y}_{s,3},\hat{y}_{s,4}\end{pmatrix}^T$ is the vector of squeezed $\hat {y}$ - quadratures used to generate cluster states. In addition, we used the rotation matrix $ R\left({\Theta} \right) $ and the squeezing matrix $S({r})$, which are defined as follows
\begin{align*} 
R\left({\Theta}\right)=\begin{pmatrix}
\cos {\Theta} & -\sin {\Theta}\\
\sin {\Theta} & \cos {\Theta}
\end{pmatrix}, \quad  S({r})=\begin{pmatrix}
\exp \left(-{r}\right) & \mathds{O}\\
\mathds{O} & \exp \left({r}\right)
\end{pmatrix}.
\end{align*} 

\begin{figure} [H]
\centering
\includegraphics[scale=1]{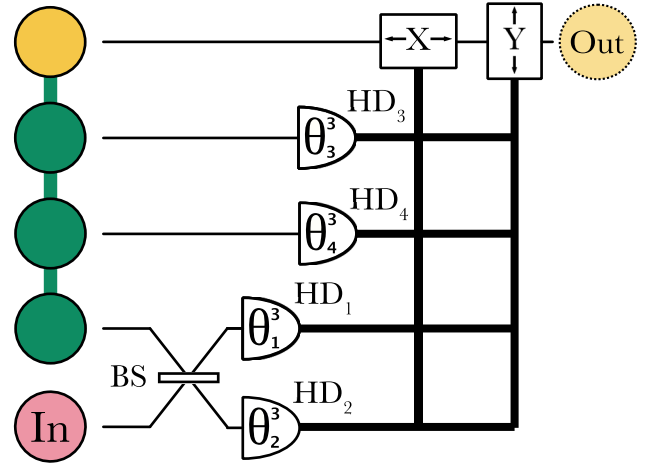}
\caption{The implementation scheme of a universal single-mode transformation using the four-node cluster state. In the figure: $\text {In}$ -- input state, $\mbox {BS}$ -- symmetric beam splitter, $\text{HD}_j$ -- homodyne detectors having local oscillators with phases $\Theta_j^3$. The dotted circle denotes the quantum state $\text {Out}$, which will result from the action of the presented scheme. In this and all other figures, $X$ and $Y$ are the designations of devices that shift the quadratures by the classical values. The bold lines on the figure indicate the classic channels through which information about the results of measurements is transmitted to shifting quadrature devices. } \label{Fig_2}
\end{figure}

Each of the five configurations shown here one can represent via a specific sequence of linear optical elements. We will discuss how to do this only for one of them - the third (see Fig. \ref{Fig_2}).
\noindent  In the presented figure, the input state with quadratures $\hat{x}_{in}$ and $\hat{y}_{in}$ mix with the external node of the linear four-node cluster state by a symmetric beam splitter. Next, the four channels are sequentially measured using homodyne detectors $\text{HD}_j$ ($j=1,\dots, 4$). Each such detector includes a local oscillator with the phase $\Theta_j^3$. The results of all measurements are sent to the devices $X$ and $Y$, which shifts the quadratures by the desired classical value. The classical contributions in the arrived quadratures can be completely eliminated using the principle of quantum teleportation. 
The resulting quadratures will be $\hat{X}_{out}$ and $\hat{Y}_{out}$. It is important to note that the theory considered by us is right for any physical system described by continuous variables. The use of optical elements in the examples is only a convention.

We have obtained that a single-mode Gaussian transform can be performed on a 4-node cluster in five different ways. The question arises whether any of these methods is preferable over others. Let us compare all the results obtained during OWQC on the clusters shown in Fig. \ref{Fig_1}. Since we have proved \cite{Korolev_2020} that all these transformations are universal single-mode transformations for $\Theta _-^j = \pi / 2$, we can only compare them by errors in the computation results. In other words, we can compare error vectors (\ref{er_1})-(\ref{er_5}) with each other.

It is correct to compare transformations that perform the same actions, i.e., convert the input state into the same output one. This can be achieved by matching the phases of the local oscillators $\lbrace \Theta_{3}^j, \Theta_{4}^j, \Theta_{+}^j \rbrace_{j=1}^5$. Then the different matrices $ {U}_{j}$ will coincide with each other. The relationships of all these phases are given by the following expressions:
\begin{align} 
&\Theta_{4}^1=\Theta_{4}^3,  \quad \Theta_{3}^2=\Theta_{3}^3, \label{phase_1}\\
&\Theta_{3}^1=\Theta_{3}^4=\Theta_{3}^5=-\arctan \left(\cot \Theta_{3}^3\right),\\
&\Theta_{4}^2=\Theta_{4}^4=\Theta_{4}^5=-\arctan \left(\cot \Theta_{4}^3\right),\\
& \Theta_{+}^1=\Theta_{+}^4=\Theta_+^3-\frac{\pi}{2}, \quad \Theta_{+}^2=\Theta_{+}^5=\pi-\Theta_{+}^3. \label{phase_4}
\end{align}
Now we can compare the errors, resulting from the same transformations in different computation schemes (Fig. \ref{Fig_1}). To this end, let us move from the error vectors (\ref{er_1})-(\ref{er_5}) to the vectors consisting of the error variance $ \langle \delta \hat{\vec{e}}_ j^2 \rangle $. In this case, we use the Eqs. (\ref{phase_1})-(\ref{phase_4}) and the fact that all squeezed oscillators involved in cluster generation are independent and have the same $\hat{y}$-quadrature variances, i.e.,
\begin{align} \label{variance}
\langle \delta \hat{y}_{s,j} \delta \hat{y}_{s,k} \rangle=\delta_{jk} \langle \delta \hat{y}_s^2\rangle,
\end{align}
where $\delta_{jk}$ is the Kronecker delta. As a result, we get equality of the form
\begin{multline}
\langle\delta \hat{\vec{e}}_1^2 \rangle=R\left(\frac{\pi}{2}\right)\langle\delta \hat{\vec{e}}_2^2 \rangle=\langle\delta \hat{\vec{e}}_3^2 \rangle\\
=R\left(\frac{\pi}{2}\right)\langle\delta \hat{\vec{e}}_4^2 \rangle=\langle\delta \hat{\vec{e}}_5^2 \rangle \equiv \langle \delta  \hat{\vec{e}}_{4modes}^2\rangle, 
\end{multline}
where 
\begin{multline} \label{error_4}
\langle \delta  \hat{\vec{e}}_{4modes}^2\rangle=\langle\delta \hat{y}_s^2 \rangle\\
*\begin{pmatrix}
2 \cot \Theta _{3} \cot \Theta_{4} \left(1 + \cot \Theta_{3} \cot \Theta_{4}\right) + 3 \csc ^2 \Theta_{4}\\
3 + 2 \cot ^2 \Theta_{3}
\end{pmatrix}. 
\end{multline}
This means that no matter what computation scheme with the four-node cluster state we choose, the error variance of the same transformations will be the same (up to the renaming of $\hat{X}$ and $ \hat{Y}$ quadratures). The result is interesting because the methods of implementing universal single-mode transformations differ from each other both in error vectors and in the graphs used.  In addition, there are infinitely many methods for generating cluster states \cite{Korolev_2018,HAL}, and, as we see, our result does not depend on them.
\subsection{CZ transformation}
\begin{figure}[H]
\centering
\includegraphics[scale=1]{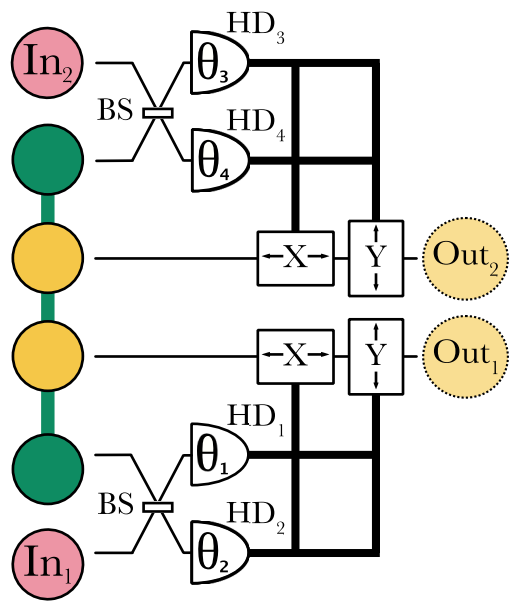}
\caption{The implementation scheme of the \emph{CZ} transformation using a four-node linear cluster state. In the figure: $\text{In}_1$ and $\text{In}_2$ are the input states; $\text{Out}_1$ and $\text{Out}_2$ are the output states that are the results of this computational scheme; $\text{HD}_j$ are the homodyne detectors that include local oscillators with phases $\Theta_2=\Theta_4=-\Theta_1=-\Theta_2=\frac{\pi}{2}$.} \label{Fig_3}
\end{figure}

We now turn to the implementation of the \emph {CZ} transformation. In \cite{Korolev_2020}, we have shown that the best case of realization this transformation (with the smallest error) is achieved when computing on cluster states with the number of nodes twice the amount of input nodes. That is, the two-mode \emph{CZ} operation requires a cluster with four nodes. Moreover, the graph of this cluster state should be linear. In Fig. \ref{Fig_3} presents an example of implementing the \emph{CZ} transformation .
\noindent  In the figure, the two input states $ \text{In}_1$ and $ \text{In}_2 $ are mixed using symmetrical beam splitters with the external cluster state nodes. Next is the measurement process. After shifting the quadratures on the devices (indicated in the diagram as $X$ and $Y$), the resulting state can be written in matrix form as follows: 
\begin{align} 
&\begin{pmatrix}
\hat{{X}}_{out,1}\\
\hat{{X}}_{out,2}\\
\hat{{Y}}_{out,1}\\
\hat{{Y}}_{out,2}
\end{pmatrix}=U_{CZ}\begin{pmatrix}
\hat{{x}}_{in,1} \\
\hat{{x}}_{in,2}\\
\hat{{y}}_{in,1}\\
\hat{{y}}_{in,2}
\end{pmatrix} +\hat{\vec{e}}_{CZ}, 
\end{align}
where a \emph{CZ} transformation matrix is given by
\begin{align}
U_{CZ}=\begin{pmatrix}
1& 0 & 0 & 0\\
0 & 1 & 0 & 0\\
0 & 1 & 1 & 0\\
1 & 0 & 0 & 1
\end{pmatrix},
\end{align} 
the error vector has the form $\hat{\vec{e}}_{CZ}=U_{CZ}\begin{pmatrix} \hat{{N}}_1 & \hat{{N}}_2 & \hat{{N}}_3 & \hat{{N}}_4 \end{pmatrix} ^T$, $\hat{N}_j$ is the nullifier of the $j$- th cluster state node. For further comparisons, we need to move from the error vector to the vector of their variances. To do this, we use the relation between nullifiers and squeezed  $\hat{\vec{y}}_s$-quadratures \cite{Korolev_2020}, as well as the Eq. (\ref{variance}). As a result, we get
\begin{align} \label{error_1}
\langle \delta \hat{\vec{e}}_{CZ}^2 \rangle =\langle \delta \hat{y}_s^2 \rangle \begin{pmatrix}
2\\
2\\
3\\
3\\
\end{pmatrix}.
\end{align}
We see that the resulting vector does not depend on the cluster state construction method. In addition, the transformation errors are independent of the local oscillator phases because certain phases have already been selected to implement this transformation.
 Note also that the considered cluster state has the minimum number of nodes necessary for implementing the \emph{CZ} transformation \cite{Korolev_2020}. It makes no sense to consider clusters with more than four nodes since they have more sources of errors (oscillators with finite squeezing) and produce larger errors. Thus, if we employ only the OWQC ideology, then the best approach will be to use ensembles of four-node cluster states. Let us now look at the implementation of universal transformations beyond the OWQC ideology. We will evaluate the errors that will be obtained in this case and compare them with the errors in the computations on cluster states discussed above.

\section{Quantum computation beyond the OWQC ideology}
So far, we have discussed which cluster configuration to choose to ensure the minimum error of the OWQC performed entirely by measuring the cluster state. We made sure that performing both single-mode and two-mode Gaussian operations requires at least four-node clusters. We got an estimate of the minimum possible computation error. In this section, we wonder whether it is possible to further reduce the computation error by going beyond pure OWQC.

\subsection{Single-mode transformations} \label{one_mode_section}
Let us start with single-mode operations. In \cite{Korolev_2020}, we have shown that a single-mode transformation performed on a two-node cluster is not universal. However, it has the form close to the desired universal one:
\begin{align} \label{2_omemode}
\begin{pmatrix} 
\hat{{X}}_{out}\\
\hat{{Y}}_{out}
\end{pmatrix}=R\left(-\frac{1}{2}{\Theta}_+^1\right)S\left(\ln\left[ \tan \frac{1}{2} {\Theta}_-^1\right]\right)\nonumber \\
*R\left(-\frac{1}{2}{\Theta}_+^1\right)\begin{pmatrix}
\hat{{x}}_{in}\\
\hat{{y}}_{in}
\end{pmatrix} + \sqrt{2}\begin{pmatrix}
\hat{y}_{s,1}\\
\hat{y}_{s,2}
\end{pmatrix}.
\end{align}
A universal single-mode Gaussian transform should be decomposed into the product of three matrices of the form $ R (\varphi_1) S (r) R (\varphi_2) $ \cite{Braunstein2005}.   Transforming the input quadratures to the output ones in the Eq. (\ref {2_omemode}) is not universal, because the rotation angles in it coincide with each other. Nevertheless, such transformation is important in terms of optimizing computation errors because it is implemented in the minimum non-trivial cluster state (the state with a minimum number of nodes, and therefore with a minimum error). Let us try to supplement this transformation up to a universal, without greatly increasing the error.

\begin{figure} [H]
\centering
\includegraphics[scale=1]{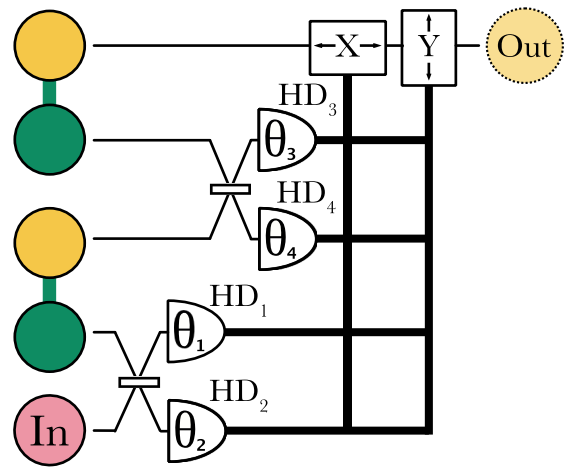}
\caption{The implementation scheme of the universal single-mode transformation using the pair of two-node cluster states. The symbols are the same as in Fig. \ref{Fig_2}.} \label{Fig_4}
\end{figure}

Consider two approaches to supplement the transform in Eq. (\ref{2_omemode}). The first approach is to use two two-node cluster state \cite{Korolev_2019,OPO}. The result of the computation in the first two-node state (quadratures (\ref{2_omemode})) is sent as input to the exactly same two-node cluster state. The implementation scheme of such a composite transformation is shown in Fig. \ref{Fig_4}
\noindent  The output quadratures obtained as a result of such computations have the following form
\begin{align} \label{One_mode}
&\begin{pmatrix} 
\hat{{X}}_{out}\\
\hat{{Y}}_{out}
\end{pmatrix} \nonumber\\
&=R\left(-\frac{1}{2}{\Theta}_+^2\right)S\left(\ln\left[ \tan \frac{1}{2} {\Theta}_-^2\right]\right)
R\left(-\frac{1}{2} {\Theta}_+^2\right) \nonumber\\
&*R\left(-\frac{1}{2}{\Theta}_+^1\right)S\left(\ln\left[ \tan \frac{1}{2} {\Theta}_-^1\right]\right)R\left(-\frac{1}{2}{\Theta}_+^1\right)\begin{pmatrix}
\hat{{x}}_{in}\\
\hat{{y}}_{in}
\end{pmatrix}  \nonumber \\
&+\sqrt{2}R\left(-\frac{1}{2}{\Theta}_+^2\right)S\left(\ln\left[ \tan \frac{1}{2} {\Theta}_-^2\right]\right)R\left(-\frac{1}{2}{\Theta}_+^2\right)\begin{pmatrix}
\hat{y}_{s,1}\\
\hat{y}_{s,2}
\end{pmatrix} \nonumber \\
&+\sqrt{2}\begin{pmatrix}
\hat{y}_{s,3}\\
\hat{y}_{s,4}
\end{pmatrix},
\end{align}
where $\Theta_{\pm}^1=\Theta_{1}\pm \Theta_2$ and $\Theta_{\pm}^2=\Theta_{3}\pm \Theta_4$ are the sums and differences of the local oscillators' phases used in homodyne detection, $\lbrace \hat{y}_{s,k}\rbrace _{k=1}^4$ are the quadratures of the squeezed quantum oscillators, which used to generate two-node cluster states.

If we put $\Theta _-^1=\frac{\pi}{2} $ in the Eq. (\ref{One_mode}), the remaining transformation will be a universal single-mode transformation of the form
\begin{multline} \label{2_1}
\begin{pmatrix} 
\hat{{X}}_{out}\\
\hat{{Y}}_{out}
\end{pmatrix}=R\left(-\frac{1}{2}{\Theta}_+^2\right)S\left(\ln\left[ \tan \frac{1}{2} {\Theta}_-^2\right]\right)\\
*R\left(-\frac{1}{2}{\Theta}_+^2-{\Theta}_+^1\right)\begin{pmatrix}
\hat{{x}}_{in}\\
\hat{{y}}_{in}
\end{pmatrix} + \hat{\vec{e}}_{2pair,1},
\end{multline}
where the error vector has the form:
\begin{multline}
\hat{\vec{e}}_{2pair,1} \\
=\sqrt{2}R\left(-\frac{1}{2}{\Theta}_+^2\right)S\left(\ln\left[ \tan \frac{1}{2} {\Theta}_-^2\right]\right)R\left(-\frac{1}{2}{\Theta}_+^2\right)\begin{pmatrix}
\hat{y}_{s,1}\\
\hat{y}_{s,2}
\end{pmatrix}\\
+\sqrt{2}\begin{pmatrix}
\hat{y}_{s,3}\\
\hat{y}_{s,4}
\end{pmatrix}.
\end{multline}
However, the transform (\ref{One_mode}) will also be universal for $\Theta_-^2=\frac{\pi}{2}$. In this case, the transformation will be given by the following equation:
\begin{multline} \label{2_2}
\begin{pmatrix} 
\hat{{X}}_{out}\\
\hat{{Y}}_{out}
\end{pmatrix}=R\left(-{\Theta}_+^2-\frac{1}{2}{\Theta}_+^1\right)S\left(\ln\left[ \tan \frac{1}{2} {\Theta}_-^1\right]\right)\\
*R\left(-\frac{1}{2}{\Theta}_+^1\right)\begin{pmatrix}
\hat{{x}}_{in}\\
\hat{{y}}_{in}
\end{pmatrix} +\hat{\vec{e}}_{2pair,2}, 
\end{multline}
where
\begin{align}
\hat{\vec{e}}_{2pair,2} =\sqrt{2}R\left(-{\Theta}_+^2\right)\begin{pmatrix}
\hat{y}_{s,1}\\
\hat{y}_{s,2}
\end{pmatrix}+ \sqrt{2}\begin{pmatrix}
\hat{y}_{s,3}\\
\hat{y}_{s,4}
\end{pmatrix}.
\end{align}

The transformations (\ref{2_1}) and (\ref{2_2}) differ from each other only by the error vector. Let us compare them by these vectors. First of all, we move from the error vectors to the vectors consisting of their variances.
\begin{align} 
&\langle \delta \hat{\vec{e}}^2_{2pairs,1} \rangle= \frac{4 \langle \delta \hat{y}_s^2\rangle}{\sin ^2  \Theta ^2_-} \begin{pmatrix}
1+\cos  \Theta_+^2\cos  \Theta_-^2\\
1-\cos  \Theta_+^2\cos  \Theta_-^2
\end{pmatrix}, \label{error_2_1}\\
&\langle \delta \hat{\vec{e}}^2_{2pairs,2} \rangle=  \langle \delta \hat{y}_s^2\rangle \begin{pmatrix}
4\\
4
\end{pmatrix}. \label{error_2_2}
\end{align}
Here, as before, we used the property (\ref{variance}). Since $\min \left[ \frac{4 \langle \delta \hat{y}_s^2\rangle}{\sin ^2  \Theta ^2_-}\left(1 \pm \cos  \Theta_+^2\cos  \Theta_-^2 \right)\right]=4 \langle \delta \hat{y}_s^2\rangle$, the error in the Eq. (\ref{error_2_1}) is always no less than the error in Eq. (\ref{error_2_2}). Thus, we have found a way to implement a universal single-mode transformation on the pair of two-node cluster states with minimal error.

Let us consider another supplement scheme for the transform (\ref{2_omemode}). To make this transformation universal, one needs to multiply it by a rotation transformation $ R\left (\varphi \right)$. For light systems, such quadrature rotation is most convenient to implement due to phase modulators placed in the light channel. If we use atomic ensembles or, for example, optomechanical systems, then it is convenient to use the free evolution of the system to perform this transformation, i. e., to leave the system to itself for a certain time. As a result of the evolution of the quadratures $\hat{ X}_{ out}$ and $\hat{ Y}_{ out}$ by the free Hamiltonian $\hat{ H}=\frac{ 1}{ 2}\left(\hat{ X}_{ out}^2+\hat{Y}_{ out}^2\right)$ , we get the desired transformation of the form:
\begin{align} \label{one-mode_rotate}
\begin{pmatrix} 
\hat{{X}}_{out}'\\
\hat{{Y}}_{out}'
\end{pmatrix}&=R\left(\varphi-\frac{1}{2}{\Theta}_+^1\right)S\left(\ln\left[ \tan \frac{1}{2} {\Theta}_-^1\right]\right)\nonumber \\
&*R\left(-\frac{1}{2}{\Theta}_+^1\right)\begin{pmatrix}
\hat{{x}}_{in}\\
\hat{{y}}_{in}
\end{pmatrix}+ \sqrt{2}R\left( \varphi \right)\begin{pmatrix}
\hat{y}_{s,1}\\
\hat{y}_{s,2}
\end{pmatrix}.
\end{align}
The vector of variances in this case has the form:
\begin{align} \label{31}
&\langle \delta \hat{\vec{e}}^2_{2oscil} \rangle=  \langle \delta \hat{y}_s^2\rangle \begin{pmatrix}
2\\
2
\end{pmatrix}.
\end{align}
Elements of the presented vector are smaller than elements of the vector $\langle \delta \hat{\vec{e}}^2_{2pairs,2} \rangle$. This result is explained by the fact that the minimum number of oscillators and the transformation that does not introduce additional errors were used to implement a universal single-mode transform in the latter case.  

Let us now compare both approaches discussed in this section with the cases of implementing universal single-mode transformations on the four-node cluster states (Section \ref{cluster_one_mode}). As before, we will compare the vectors of variances. Let us compare at first  the vector
\begin{align} \label{err_44}
&\langle \delta  \hat{\vec{e}}_{4modes}^2\rangle=\langle\delta \hat{y}_s^2 \rangle \nonumber\\
&*\begin{pmatrix}
2 \cot \Theta _{3} \cot \Theta_{4} \left(1 + \cot \Theta_{3} \cot \Theta_{4}\right) + 3 \csc ^2 \Theta_{4}\\
3 + 2 \cot ^2 \Theta_{3}
\end{pmatrix}, 
\end{align}
with the vector $\langle \delta  \hat{\vec{e}}_{2oscil}^2\rangle$, given by Eq. (\ref{31}). Comparison will be done componentwise. Since in this work we look for configurations of cluster states that give the smallest error, we need to begin the comparison by searching for the minimum values of the components of the vectors. For the vector $\langle \delta \hat{\vec{e}}_{4modes}^2\rangle$, the minimum values of the components coincide and are equal to 
\begin{align*}
&\min \limits_{\Theta_3,\Theta_4} \lbrace 2 \cot \Theta _{3} \cot \Theta_{4} \left(1 + \cot \Theta_{3} \cot \Theta_{4}\right) + 3 \csc ^2 \Theta_{4}\rbrace\langle \delta \hat{y}_s^2\rangle\\
&=\min \limits_{\Theta_3} \lbrace 3 + 2 \cot ^2 \Theta_{3} \rbrace\langle \delta \hat{y}_s^2\rangle= 3\langle \delta \hat{y}_s^2\rangle.
\end{align*}
This value is greater than $2\langle \delta \hat{y}_s^2\rangle$ (always greater than the value of the x components).  It follows that the error of the universal single-mode transformations implemented on four-node cluster states is always greater than the error obtained by using a single two-node state and the additional quadrature rotator.

\begin{figure*}
\includegraphics[scale=0.72]{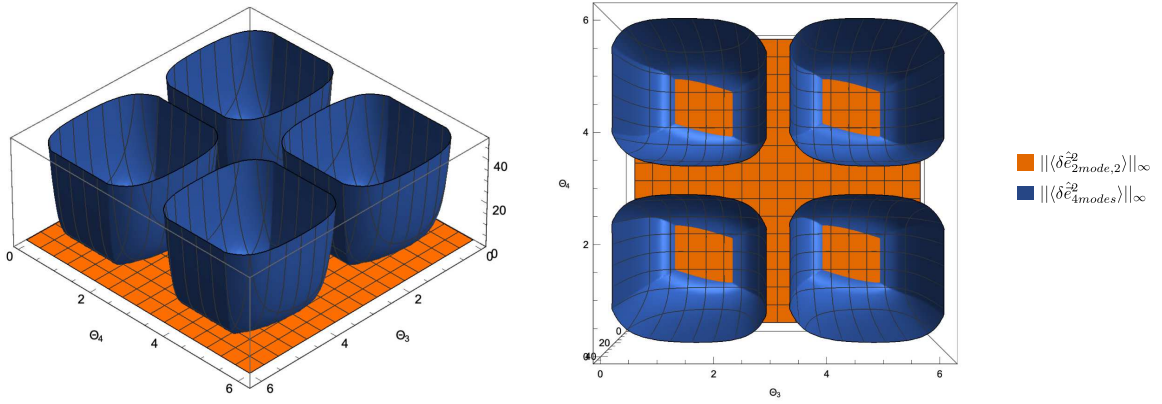}
\caption{Three-dimensional surfaces of errors in two computation schemes. The figure shows two projections of the same surfaces. The blue color in the graph indicates the distribution of errors $||\langle \delta \vec{\hat{e}}_{4modes}^2\rangle||_{\infty}$, and  orange -- $||\langle \delta \vec{\hat{e}}_{2pairs,2}^2\rangle||_{\infty}$.} \label{Fig_5}
\end{figure*}

To complete the picture, let us also compare the vector of error variances obtained when computing on four-node cluster states (\ref{err_44}) with the vector of error variances in the best computation case on pairs of two-node cluster states (\ref{error_2_2}). As we have already clarified, the minimum value of the elements of the vector (\ref{err_44}) is $3\langle\delta\hat {y}_s^2\rangle$. This is less than the elements of the (\ref{error_2_2}) vector. This means that the computation scheme with the pair of two-node cluster states gives a large error for some transformations. However, the situation is the opposite for other transformations. Moreover, for some transformations, the error variances in the scheme with a four-node cluster is greater only by one quadrature, and for some by two at once. The above means that it is not evident which computation case is better. Nevertheless, we can compare the two approaches by the number of transformations in which the result has a smaller error for two quadratures. To this end, we move from the vectors of error variances (\ref{error_2_2}) and (\ref{err_44}) to their $L_{\infty}$ norms, which are defined by the equation $||\vec v||_{\infty}=\max \limits_{i} |v_i|$. Using these norms, it is convenient to compare the maximum errors in quadratures for different values of the parameters $\Theta_3$ and $\Theta_4$. For the considered vectors of error variances, these norms are given by the following equations:
\begin{align}
&||\langle \delta  \hat{\vec{e}}^2_{4modes}\rangle ||_{\infty}\nonumber\\
&= \max \limits_{ \Theta_3, \Theta_4 } \lbrace  \cot \Theta _{3} \cot \Theta_{4} \left(1 + \cot \Theta_{3} \cot \Theta_{4}\right) + 3 \csc ^2 \Theta_{4} ; \nonumber \\
&3 + 2 \cot ^2 \Theta_{3}  \rbrace \langle\delta \hat{y}_s^2 \rangle  \label{delta_1}\\
&||\langle \delta  \hat{\vec{e}}_{2mode,2}^2\rangle ||_{\infty}= 4\langle\delta \hat{y}_s^2 \rangle
\end{align}
Since the value $\langle\delta \hat{y}_s^2 \rangle$ is the same in both equations, we can only compare the coefficients for it. For clarity of comparison, it is convenient to plot functions $||\langle \delta  \hat{\vec{e}}_{4modes}^2\rangle ||_{\infty}$ and $|\langle \delta  \hat{\vec{e}}_{2mode,2}^2\rangle ||_{\infty}$ in the coordinates $\left( \Theta_{3}, \Theta_{4}\right)$. The graph of these functions is shown in Fig. \ref{Fig_5}.
\noindent This graph shows the errors of different transformations in two computation schemes. The surface of the function $||\langle\delta\hat{\vec{e}}_{4modes}^2\rangle||_{ \infty}$ in the coordinates $\Theta_3$, $\Theta_4$ resembles the shape of a yogurt cups: in the four areas corresponding to the bottoms of the cups, the error vector norm is minimal. This norm increases quite sharply, imitating the walls of the yogurt cups.

From Fig. \ref {Fig_5} one can see in which areas the computation on four-node cluster states give a smaller error, and in which the situation is opposite. To compare the two approaches by the number of transformations with a smaller error, we find the area $S_1$ occupied by points $\Theta_{3}$ and $\Theta_{4}$, in which  $||\langle \delta \vec{\hat{e}}_{4modes}^2\rangle ||_{\infty}<||\langle \delta \vec{\hat{e}}_{2pairs,2}^2\rangle ||_{\infty}$, and  area $S_2$, where  $||\langle \delta \vec{\hat{e}}_{2pairs,2}^2\rangle||_{\infty} < ||\langle \delta \vec{\hat{e}}_{4modes}^2\rangle||_{\infty}$. Taking the ratio of these two areas, we get
\begin{align*}
\frac{S_2}{S_1}\approx 6.
\end{align*}
In other words, computations with the pair of two-node cluster states give a smaller error for a larger number of transformations. Moreover, when implementing stretch transformations in the computational scheme with the four-node cluster state (for $\Theta_ {3} $ and $ \Theta_ {4} $ close to $ \pi$), the maximum error will increase proportionally to the stretching, as shown in Fig.  \ref{Fig_5}.  In the computation scheme with the pair of two-node cluster states, this will not happen because the error there is always the same.

In sum, we can conclude that the best case for implementing universal single-mode Gaussian transformations is obtained when computing on the two-node cluster state with an additional quadrature rotator. Next, in terms of the error value, is the case of computations on the pairs of two-node cluster states. As it turned out, the computation case on a four-node cluster state has the worst error in the results.

\subsection{CZ transformation}
Let us consider other schemes for implementing the \emph{CZ} transformation. As before, we will leave the usual OWQC ideology. We will compare all possible configurations with each other.

\begin{figure}[H]
\centering
\includegraphics[scale=0.9]{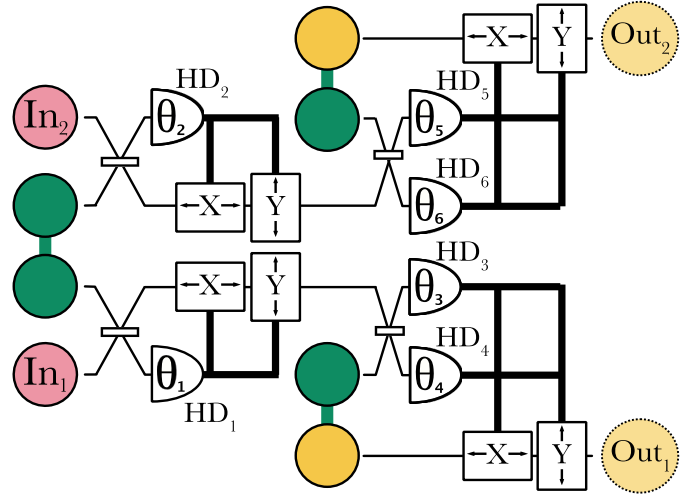}
\caption{The implementation scheme of the \emph{CZ} transformation using the two-node cluster states. In the figure: $\text{In}_1$, $\text{In}_2$ are the input states, $\text{Out}_1$ and $\text{Out}_2$ are the output states. Here the phases of the homodyne detectors are $\Theta_2=\Theta_1=0$ and $\Theta_6=\Theta_3=-\Theta_4=-\Theta_5=-\arctan \left[\ln 2/2\right]$. The other symbols are the same as in Fig. \ref{Fig_2}.} \label{Fig_6}
\end{figure}

As the first example, we consider the implementation of \emph{CZ} transformation on two-node cluster states. In \cite{Korolev_2020}, we proved that it is possible to implement an operation of type \emph{CZ} using two-node cluster states. This transformation is given by the following equation
\begin{align} \label{comp_CZ_1}
\begin{pmatrix}
\hat{{X}}_{out,1}' \\
\hat{{X}}_{out,2}'\\ 
\hat{{Y}}_{out,1}'\\
\hat{{Y}}_{out,2}' 
\end{pmatrix}=\begin{pmatrix}
\sqrt{2} & 0 & 0 & 0\\
0 & \sqrt{2} & 0 & 0\\
0 & 0 & \frac{1}{\sqrt{2}} & 0\\
0 & 0 & 0 & \frac{1}{\sqrt{2}}
\end{pmatrix}  U_{CZ}\begin{pmatrix}
\hat{{x}}_{in,1} \\
\hat{{x}}_{in,2}\\
\hat{{y}}_{in,1}\\
\hat{{y}}_{in,2}
\end{pmatrix}-\begin{pmatrix}
0\\
0\\
\hat{y}_{s,1}\\
\hat{y}_{s,2}
\end{pmatrix}.
\end{align}
As you can see, we do not get the pure \emph{CZ} transformation. We have a joint action of the \emph{CZ} operator and the $\hat{y}$- quadrature squeezing operators. If we supplement these transformations by stretching the two output $\hat {y}$-quadratures, we get the pure \emph{CZ} transformation. From the Eq. (\ref{2_omemode}) we see that single-mode stretching can be implemented using a two-node cluster state (when ${\Theta} _- = -2\arctan \left (\frac {\ln 2} {2} \right) $ and $ {\Theta _ +} = 0 $). To stretch two $\hat{y}$-quadratures, one should use a pair of two-node clusters. The result of this stretching can be written in vector form as follows:
\begin{align} \label{s_1}
\begin{pmatrix}
\hat{{X}}_{out,1} \\
\hat{{X}}_{out,2}\\
\hat{{Y}}_{out,1} \\
\hat{{Y}}_{out,2}
\end{pmatrix}=\begin{pmatrix}
\frac{1}{\sqrt{2}} & 0\ & 0 & 0\\
0  & \frac{1}{\sqrt{2}}& 0 & 0\\
0 & 0 & \sqrt{2} & 0\\
0 & 0& 0 & \sqrt{2}
\end{pmatrix} \begin{pmatrix}
\hat{{x}}_{in,1}'  \\
\hat{{x}}_{in,2}'\\
\hat{{y}}_{in,1}'\\
\hat{{y}}_{in,2}'
\end{pmatrix}+\sqrt{2}\begin{pmatrix}
\hat{y}_{s,3}\\
\hat{y}_{s,4}\\
\hat{y}_{s,5}\\
\hat{y}_{s,6}
\end{pmatrix}.
\end{align}
By applying this transformation to the quadratures from the Eq. (\ref{comp_CZ_1}) (substituting quadratures (\ref {comp_CZ_1}) as the input ones to the Eq. (\ref {s_1})), we get the pure \emph{CZ} transformation of the form:
\begin{align} \label{qvad_add}
\begin{pmatrix}
\hat{{X}}_{out,1} \\
\hat{{X}}_{out,2}\\ 
\hat{{Y}}_{out,1}\\
\hat{{Y}}_{out,2} 
\end{pmatrix}=U_{CZ}\begin{pmatrix}
\hat{{x}}_{in,1} \\
\hat{{x}}_{in,2}\\
\hat{{y}}_{in,1}\\
\hat{{y}}_{in,2}
\end{pmatrix}+\sqrt{2}\begin{pmatrix}
 \hat{y}_{s,3}\\
 \hat{y}_{s,4}\\
\hat{y}_{s,5}-\hat{y}_{s,1}\\
\hat{y}_{s,6}- \hat{y}_{s,2}
\end{pmatrix},
\end{align}
where $\lbrace \hat{y}_{s,j}\rbrace_{j=1}^6$  are the squeezed quadratures of oscillators used in the computation. It is important to note that in implementing this transformation, we added two-node cluster states to the computational scheme. This, in turn, introduced an additional error in the results (the squeezed quadratures $\lbrace\hat {y} _ {s, i}\rbrace_{i=3}^6$ in the Eq. (\ref{qvad_add})). In Fig. \ref{Fig_6} presents an example of implementing the \emph{CZ} transformation using the described method.
\noindent  The first part of the scheme implements the transformation (\ref{comp_CZ_1}). Further, the result of this transformation is sent to the input of two stretch transformations (\ref{s_1}), realized using two-node cluster states.

To compare the resulting transformation with the other, we find its vector of error variances.
\begin{align*}
\langle \delta \hat{\vec{e}}_{CZ,1}^2 \rangle =\langle \delta \hat{y}_s^2 \rangle \begin{pmatrix}
2\\
2\\
4\\
4\\
\end{pmatrix}, 
\end{align*}
where we used the property (\ref{variance}). Comparing this vector with the vector (\ref{error_1}) obtained in computation on the four-node cluster state, we see that  $||\langle \delta \hat{\vec{e}}_{CZ}^2 \rangle||_{\infty}<||\langle \delta \hat{\vec{e}}_{CZ,1}^2 \rangle||_{\infty}$. In other words, the error of computations on the four-node cluster state is fewer.

\begin{figure}[H]
\includegraphics[scale=1.2]{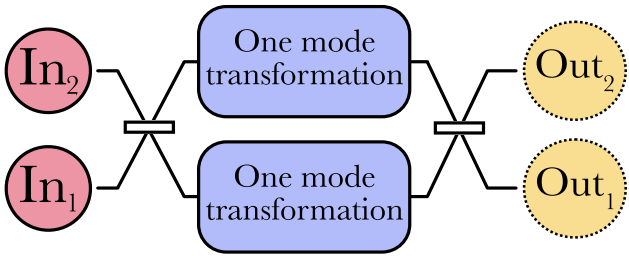}
\centering
\caption{The implementation scheme of the \emph{CZ} transformation.} \label{Fig_7}
\end{figure}

Let us now consider another scheme for implementing the CZ \cite{Korolev_2019}  transformation. This scheme is shown in Fig. \ref{Fig_7}.
\noindent In this scheme, two input states are mixed on symmetrical beam splitters. Further, independent single-mode universal transformations are applied to each of them. After that, the states obtained after the transformations are again mixed on a symmetric beam splitter. We believe that ideal beam splitters do not introduce additional errors into the computation results. The main error source here is single-mode transformations.

To implement the \emph{CZ} transformation in this scheme, one needs to perform the two single-mode transformations of the form
\begin{align} \label{AB}
A=\begin{pmatrix}
1 & 0 \\
1 & 1
\end{pmatrix}, \qquad B=\begin{pmatrix}
1 & 0 \\
-1 & 1
\end{pmatrix}.
\end{align}
As we have already known, single-mode transformations are best implemented using a two-node cluster state and additional quadrature rotation (section \ref{one_mode_section}). To implement the $A$ transform, one need to substitute $\varphi=\pi/2$, $\Theta_-^1=\arctan 2$, $\Theta_+^1=\arctan 2$ in the Eq. (\ref{one-mode_rotate}). For implementing the $B$ transformation, we need the following equalities: $\varphi=\pi/2$, $\Theta_ - ^1= - \arctan 2$, $\Theta_+^1=-\arctan 2$. As a result, errors in each of these transformations are given by the following equations:
\begin{align*}
\vec{\hat{e}}_{A}= \sqrt{2}\begin{pmatrix}
\hat{y}_{s,2}\\
-\hat{y}_{s,1}
\end{pmatrix}, \quad \vec{\hat{e}}_{B}=\sqrt{2}\begin{pmatrix}
\hat{y}_{s,4}\\
-\hat{y}_{s,3}
\end{pmatrix}.
\end{align*}
After the second beam splitter, the errors are mixed. As a result, the general error vector obtained in this implementation of the \emph{CZ} transformation has the following form
\begin{align*}
\vec{\hat{e}}_{CZ,2}=\begin{pmatrix}
\hat{y}_{s,2}+\hat{y}_{s,4}\\
\hat{y}_{s,2}-\hat{y}_{s,4}\\
-\hat{y}_{s,1}-\hat{y}_{s,3}\\
-\hat{y}_{s,1}+\hat{y}_{s,3}
\end{pmatrix},
\end{align*}
Moving to the vector of variances, taking into account Eq. (\ref{variance}), we get
\begin{align*}
\langle \delta \hat{\vec{e}}_{CZ,2}^2 \rangle = \langle \delta \hat{y}_{s}^2\ \rangle\begin{pmatrix}
2\\
2\\
2\\
2
\end{pmatrix}.
\end{align*}
We see that in this case $||\langle \delta \hat{\vec{e}}_{CZ,2}^2 \rangle||_{\infty} \textless||\langle \delta \hat{\vec{e}}_{CZ}^2 \rangle||_{\infty} \textless ||\langle \delta \hat{\vec{e}}_{CZ,1}^2 \rangle||_{\infty}$. This means that the error here has the lowest value among all the cases considered. Moreover, since the single-mode transformations used in this scheme have minimal error, consideration of other implementations of single-mode transformations will give the worst result.

\begin{figure*}
\centering
\includegraphics[scale=0.8]{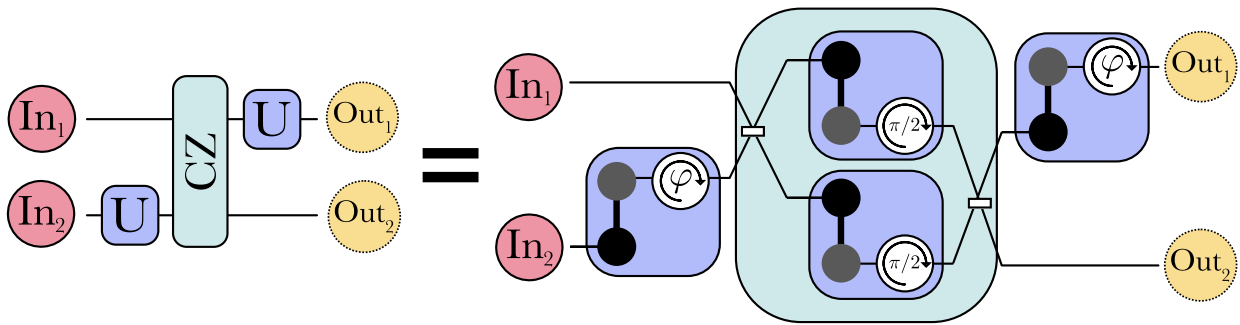}
\caption{The figure shows the strategy for constructing the quantum transformation with minimal error. The selected transformation (left) and the corresponding optimal implementation (right) are presented. As an example, we considered a transformation over two input states ($\text{In}_1$ and $\text{In}_2$), consisting of two single-mode transformations U and one \emph{CZ} transform. The computation is implemented using two-node cluster states and additional phase rotators.} \label{Fig_8}
\end{figure*}

Thus,  it is best to use two-node cluster states with additional quadrature rotators to implement both universal single-mode and \emph{CZ} transformations. Each such elementary transforms is performed with a minimum error therefore the error in the complete transformation will also be minimal.  Fig. \ref {Fig_8} shows an example of the physical implementation of an arbitrary computation scheme using two-node cluster states.

\section{Conclusion}
In this work, we investigated approaches to optimal Gaussian computations on cluster states of various configurations. We analyzed the errors that are obtained in implementing universal single-mode transformations and the two-mode transform \emph{CZ}.  At the same time, we did not confine ourselves to the traditional one-way quantum computations, when the desired operation is realized on one cluster state. We have considered cluster states with small node number, which, together with auxiliary elements, can give universal computation.

As a result, we obtained that the hybrid computing method, which uses cluster states with a small number of nodes and additional linear elements, allows us to perform the entire set of required Gaussian transformations. Moreover, this method significantly reduces the computation error, compared to the traditional OWQC approach.

As a result, we obtained that the hybrid computing method, which uses cluster states with a small number of nodes and additional linear elements, allows us to perform the entire set of required Gaussian transformations. Moreover, this method significantly reduces the computation error, compared to the traditional OWQC approach. The best result is obtained by computation on two-node cluster states supplemented by a quadrature rotator. A quadrature rotator can be implemented experimentally, for example, using a phase modulator placed in the light channel. Moreover, in implementing the \emph{CZ} transformation, it is not necessary to rotate the quadratures by an arbitrary angle, it is enough to place the glass plate in each channel, which will rotate quadratures by the angle $\pi/2$.

In this way,  the Gaussian computations with a minimal error are realized, when for all single-mode transformations we use a two-node cluster state with an additional rotator, and the two-mode operation \emph {CZ}  is performed using the scheme shown in Fig. \ref {Fig_4}.  Since, in this case, the error is minimal, it is easiest to compensate it with quantum error correction codes.

\section{Acknowledgment}
This work was supported by the Russian Foundation for Basic Research (Grant Nos 18-32-00255mol\_a, 19-02-00204a and 18-02-00648a) and as part of the research activities of the Center of Quantum Technologies, Moscow State University M.V. Lomonosov.

\end{document}